\documentclass[twocolumn,twoside,slac_two]{revtex4}
\usepackage{graphicx}
\usepackage{fancyhdr}

\usepackage{amsmath}

\newcommand{\X}[2][]{\mbox{$\tilde{\chi}_{#2}^{#1}$}}
\newcommand{\sG}[1][]{\mbox{$\tilde{g}_{#1}$}}
\newcommand{\sQ}[1][] {\mbox{$\tilde{q}_{#1}$}}

\newcommand{\sTp}[1][]{\mbox{$\tilde{t}_{#1}$}}
\newcommand{\sBt}[1][]{\mbox{$\tilde{b}_{#1}$}}

\newcommand{\ET}{E_T}
\newcommand{\MET}{\rm{E}_T^{miss}}
\newcommand{\sTau} {\mbox{$\tilde{\tau_{1}}$}}
\newcommand{\chipm} {\mbox{$\tilde{\chi}_{1}^{\pm}$}}
\begin{document}

%Title of paper
\title{Inclusive search for SUSY in top final states at CMS}

% Repeat the \author .. \affiliation  etc. as needed
%
% \affiliation command applies to all authors since the last
% \affiliation command. The \affiliation command should follow the
% other information

\author{Saeid Paktinat Mehdiabadi}
\affiliation{IPM \& Sharif University of Technology, Tehran, Iran}
%
%\author{P. Lucas}
%\affiliation{FNAL, Batavia, IL 60510, USA}

\begin{abstract}
An inclusive search for low mass SUSY using the top plus missing
transverse energy signature at CMS is presented. A 5$\sigma$
excess can be observed with $\sim$250 pb$^{-1}$ at a particular
low mass SUSY point. With the  same analysis selection the
5$\sigma$ discovery reach contours in the mSUGRA parameter space
for 1 and 10 fb$^{-1}$ are obtained.
\end{abstract}

%\maketitle must follow title, authors, abstract
\maketitle

\thispagestyle{fancy}

% body of paper here - Use proper section commands
% References should be done using the \cite, \ref, and \label commands
% Put \label in argument of \section for cross-referencing
%\section{\label{}}

\section{Introduction}
Although the standard model (SM) of the elementary particles
explains precisely the data from the colliders, there are some
remaining fundamental questions raised by the SM itself that
don't have any explanation.
\begin{figure}[!Hhtb]
\begin{center}
\resizebox{6cm}{!}{\includegraphics[70,130][300,190]{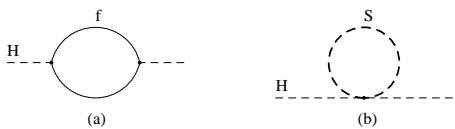}}
\caption{Loops affecting the squared Higgs mass from (a) fermions
trilinear couplings, (b) scalars quartic couplings
\cite{Martin97}.} \label{fig.mssm.divergences.loops}
\end{center}
\end{figure}
For example radiative loop corrections to the higgs mass like in
figure \ref{fig.mssm.divergences.loops} give a correction
\begin{equation}
 \Delta m_H^2|_f = \frac{\lambda_f^2}{8\pi^2} [- \Lambda^2
              + 6 m_f^2 ln \frac{\Lambda}{m_f} ]
\end{equation}
where $\Lambda$ is the upper limit of the momentum. The leading
term diverges quadratically (The {\bf Quadratic Divergency
Problem}). This divergence can be removed by introducing
 a cut-off scale on momentum. Its value is the higher limit of energy upto which SM
 is valid. Usually this limit is $M_{Pl} \equiv (8 \pi G_{Newton})^{-1/2} = 2.4 \times
10^{18}$ GeV, at which gravity has a strength comparable to the
other interactions and SM must be modified. This enormous
disparity of scales between electroweak and $M_{Pl}$ is not
natural and is called the {\bf hierarchy problem} \cite{Martin97}.
If another scalar exists which couples to the Higgs by a quartic
interaction of the form $-\lambda_S |H|^2 |S|^2$ (see figure
\ref{fig.mssm.divergences.loops}(b)), it contributes to the Higgs
mass by:
\begin{eqnarray}
 \Delta m_H^2|_S = \frac{\lambda_S}{16\pi^2} [\Lambda^2
              - 2 m_S^2 ln \frac{\Lambda}{m_S} ]
\label{eq.mssm.divergences.boson}
\end{eqnarray}
It also generates a quadratic divergence, but because of Fermi
statistics, its sign is opposite to the one of the fermions.
Assuming there are two scalar partners for every fermion with
couplings $\lambda_S = \lambda_f^2$ the quadratic divergences
cancel exactly.

To solve the hierarchy problem this theory should contain nearly
degenerate fermions and scalars and adequately chosen couplings.
It is exactly what
 {\bf supersymmetry} (SUSY) postulates, the existence of supersymmetric
partners for every SM particle, which have exactly the same
quantum numbers and mass, but differ by 1/2 in their spin.

The super partners of the fermions are called with same name
starting with ``s'', e.g stop, and the superpartners of bosons
are called with same name ending with ``ino'', e.g wino.

SUSY particles have not been discovered yet, so they are not at
the same mass as their SM partners. It means that supersymmetry
is a broken symmetry and SUSY particles are heavier. There are
different mechanisms to break this symmetry softly (without
generating quadrative divergencies). Here we consider a very
constrained scenario where gravity is responsible for SUSY
breaking and the $Z^0$ mass is produced by radiative electroweak
symmetry breaking. This model is called {\bf mSUGRA}. The
particle masses and branching ratios in mSUGRA are defined
completely by only 5 parameters:
\begin{description}
\item[$m_0$] commom scalar mass at the Grand Unification Theories (GUT) scale.
\item[$m_{1/2}$] commom gaugino mass at GUT scale.
\item[$A_{0}$] commom trillinear coupling at GUT scale.
\item[tan$\beta$] ratio of the vacuum expectation values for $H_u$ and $H_d$.
\item[sign($\mu$)] $\mu$ is the higgs mixing parameter.
\end{description}

In the following the search for SUSY in a special channel in
 CMS \cite{CMS} at the Large Hadron
Collider (LHC) \cite{LHC} which is going to start data taking in
fall 2007 will be reviewed by one example emphasizing the role of
the top quark as both signal and background.

%\paragraph{\large{CMS}}

\section{Cut Optimization in a CMS Test Point}
\begin{figure}[!Hhbt]
\begin{center}
%  \resizebox{6cm}{!}{\includegraphics{/localscratch/p/paktinat/analysis/TopSusy/Chi2NoCutScaled1.eps}}
  \resizebox{6cm}{!}{\includegraphics[0,0][550,420]{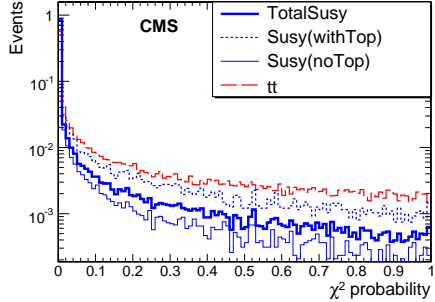}}
\caption{$\chi^2$ probability distribution for 200K events of the
inclusive
  LM1 sample and 100K events of the inclusive $t\bar{t}$ sample. All plots are
  normalized to 1 to make the shape comparison easier. The fake top quarks (SUSY events without a generated top
  quark `Susy(noTop)')
  are concentrated in the low probability region.}
  \label{fig:Chi2KinFit}
\end{center}
\end{figure}

\begin{figure}[!Hhtb]
 \begin{center}
\resizebox{4cm}{!}{\includegraphics[0,0][550,420]{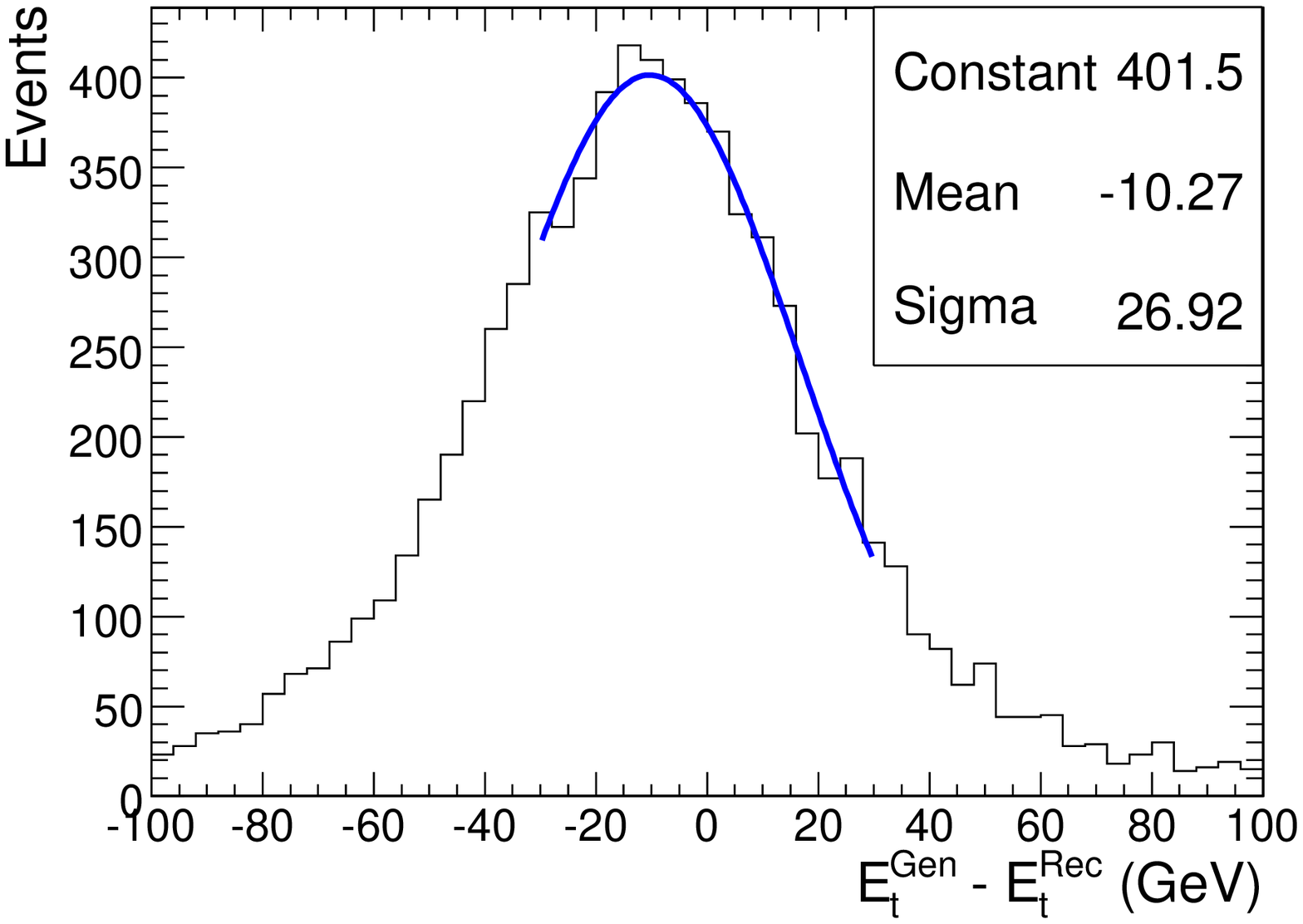}}
\resizebox{4cm}{!}{\includegraphics[0,0][550,420]{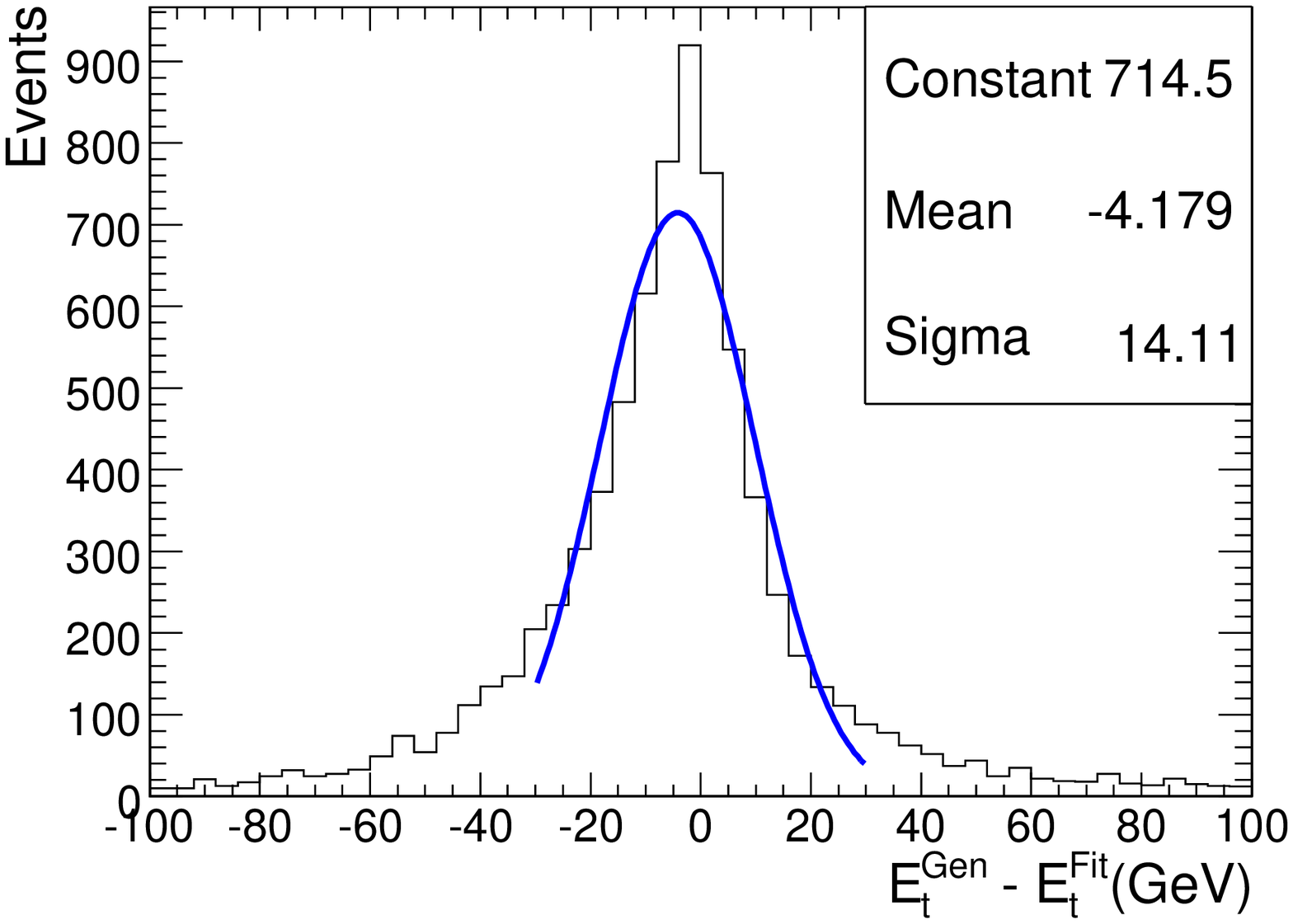}}
\caption{The difference between the energy of the
  reconstructed/fitted top and the generated top. Fitted jet combinations pass the
  probability cut($\chi^2$ probability $>0.05$). The central
  parts of the distributions(-30,30) are fitted with a gaussian function(thick-blue
  lines) to emphasize and quantify the improvement in the resolution. The fit
  improves the resolution of the energy of the W and top quark by 37\% and 48\%,
  respectively. It also improves the bias. Distributions are made by
  100k events of the inclusive
  $t\bar{t}$ sample.}
  \label{fig.paktinat.KinFitImpr}
\end{center}
\end{figure}

In CMS an inclusive search for SUSY is done by looking for events
with a top quark in the final states \cite{PaktinatTopSusy}. For
illustration a test point LM1 within the mSUGRA scenario is used.
This point is defined by $m_0 = 60$, $m_{1/2}=250$, $A_0 = 0$,
$tan\beta = 10$ and $\mu > 0$. The masses of the relevant
particles (in GeV/$c^2$) are
\begin{eqnarray}
%m(\sG) &=611 &\qquad     m(\sTp[1]) &= 412 &\qquad m(\sTp[2]) &=
%576\\\nonumber m(\X[0]{1}) &= 120 &\qquad m(\sBt[1]) &= 514
%&\qquad m(\sBt[2]) &= 535
m(\sG) =611 &    m(\sTp[1]) = 412 & m(\sTp[2]) = 576\\\nonumber
m(\X[0]{1}) = 120 & m(\sBt[1]) = 514 & m(\sBt[2]) = 535
\end{eqnarray}
In this point, the top quark can be produced inclusively in the
decay of the $\sTp$, $\sBt$ and $\sG$ accompanied by a neutralino
(the lightest supersymmetric particle (LSP)), which appears as a
large missing transverse energy (MET, $\MET$).  %Hence the SUSY
%events have a large $\MET$,
So the idea is to see the excess in the number of the extracted
top quarks, when a hard cut is applied on the missing transverse
energy. To simulate the detector response, the full simulation
based on GEANT4, OSCAR \cite{OSCAR}, is used . To extract the top
quark, a two constraints kinematic fit is utilized
\cite{KinFitAN2005025}. It is very useful because firstly it has
a quantitative feature to reject the fake top quarks ($\chi^2$
probability, see figure \ref{fig:Chi2KinFit})
%shows the $\chi^2$ probability distribution for 100k events of
%the $t\bar{t}$ sample and 200k events of the LM1 sample. All plots are
%normalized to 1, to make the shape comparison easier.
and also it can improve the kinematic features of the
reconstructed top quark. Figure \ref{fig.paktinat.KinFitImpr}
shows the difference between the energy of the reconstructed
(fitted) top quark and the generated top quark, when the generated
top quark which is closer than $\Delta R = \sqrt{\Delta \eta^2 +
\Delta \phi^2} = 0.5$ to the reconstructed top quark
%a generated top quark that decays
%hadronically and all of its partons pass the kinematic cuts of the
%jets($E_t \ge 30$ GeV and $|\eta| \le 2.5$)}
decays hadronically and all its three partons have $E_T >$ 30 GeV
and $|\eta| <$ 2.5.

The main background is semileptonic  $t\bar{t}$ when there is a
potentially high value for $E_T^{miss}$ and a hadronically
decaying top quark. In these events $E_T^{miss}$ and the fitted
top quark are almost back-to-back in the transverse plane. This
feature is used to suppress the $t\bar{t}$ background. The cuts
were optimized both to suppress the SM background and also
increase the ratio of the SUSY events with a generated top quark
`SUSY(withTop)' against the SUSY events without a generated top
quark `SUSY(noTop)', although these two sources are used as the
number of extracted signal events. In order to do the mentioned
suppressions the following cuts are applied:
\begin{enumerate}
\item The first level trigger (L1)  {\small L1JetMET}
 %\cite{HLT-TDR}
\cite{PTDR-V1}
 (a barrel jet with $\ET > 88$ and
$\MET > 46$ GeV).
\item The High Level Trigger
(HLT) {\small HLTJetMET} (a barrel jet with $\ET > 180$ and $\MET
> 123$ GeV).
\item  $\MET$, $>$ 150GeV, as shown in figure
\ref{fig.paktinat.METAfetJetCut}.
\begin{figure}[!Hhtb]
% \begin{center}
\resizebox{6cm}{!}{\includegraphics{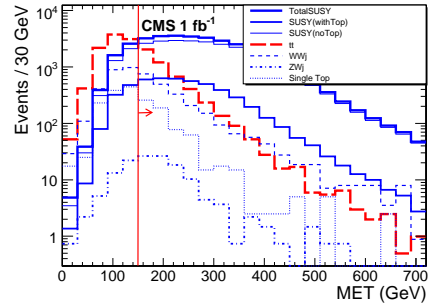}}
\caption{$\MET$ distributions for different samples. Events are
required to pass the Level 1 and High Level Trigger.}
\label{fig.paktinat.METAfetJetCut}
%\end{center}
\end{figure}
\item at least 4 jets, with at least one of them b-tagged, jets with $E_T^{uncorrected}
>$ 30 GeV and $|\eta| <$ 2.5.
\item A convergent fit with $\chi^2$ probability $>$ 0.1. This cut
is the most important cut to increase the ratio of SUSY(withTop)
against the SUSY(noTop).
\item $\Delta \phi$ between the fitted top quark and the $E_T^{miss}
<$ 2.6. Figure \ref{fig.paktinat.DeltaPhi}
\begin{figure}[!Hhtb]
% \begin{center}
\resizebox{6cm}{!}{\includegraphics{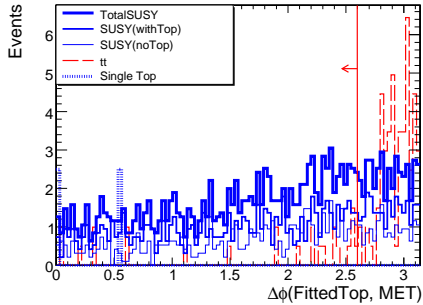}}
\caption{The $\Delta \phi$ between the fitted top and the missing
transverse energy. The distributions are made after applying the
cut 5.} \label{fig.paktinat.DeltaPhi}
%\end{center}
\end{figure}
 shows the distribution
of this quantity for different samples.
\item at least one isolated electron or muon with $P_T >$ 5 GeV/$c$
and $|\eta| <$ 2.5. This cut is introduced to suppress the QCD
multijet backgrounds.
\end{enumerate}

\begin{figure}[!Hhtb]
 \begin{center}
\resizebox{4cm}{!}{\includegraphics{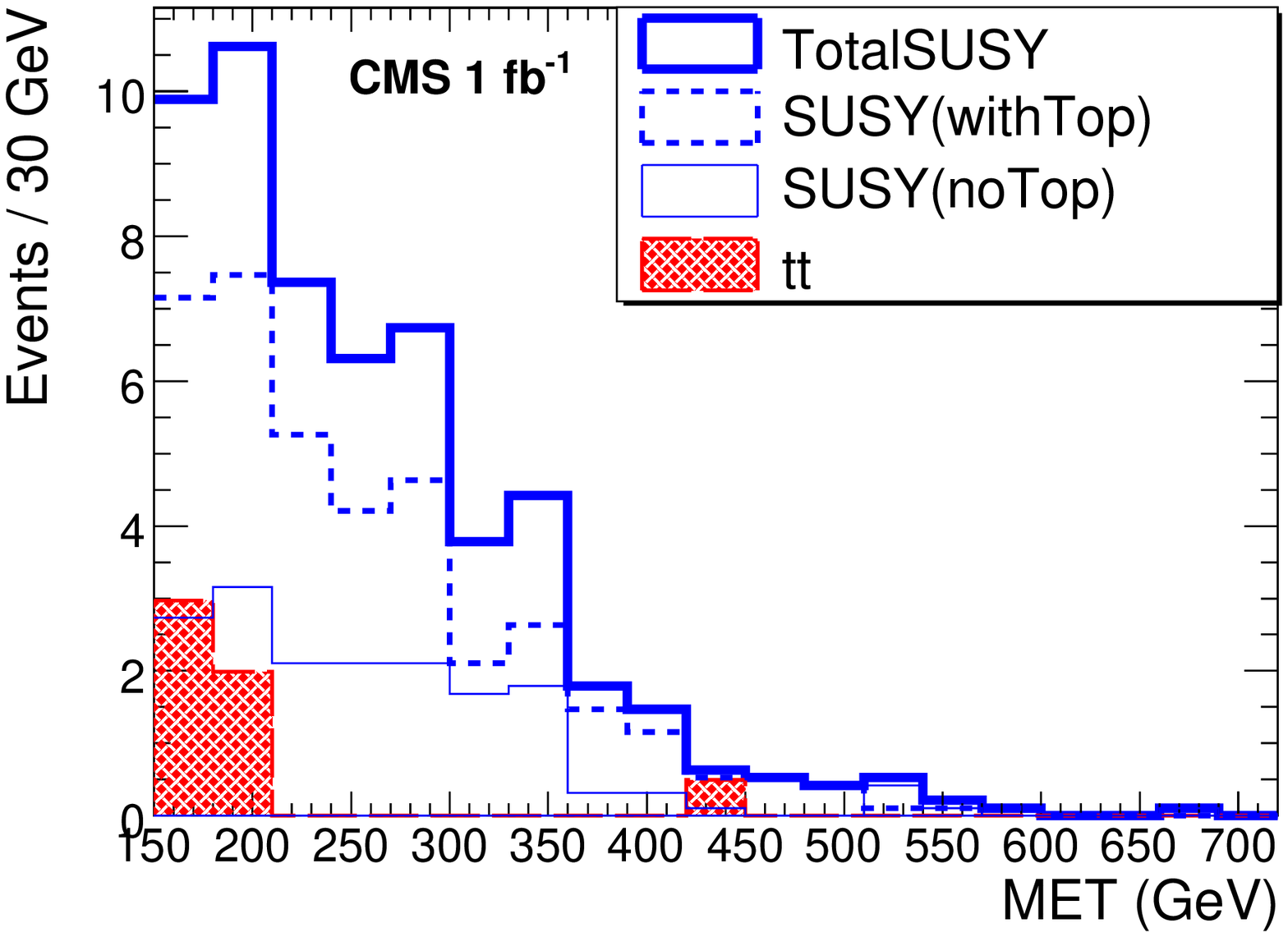}}
\resizebox{4cm}{!}{\includegraphics{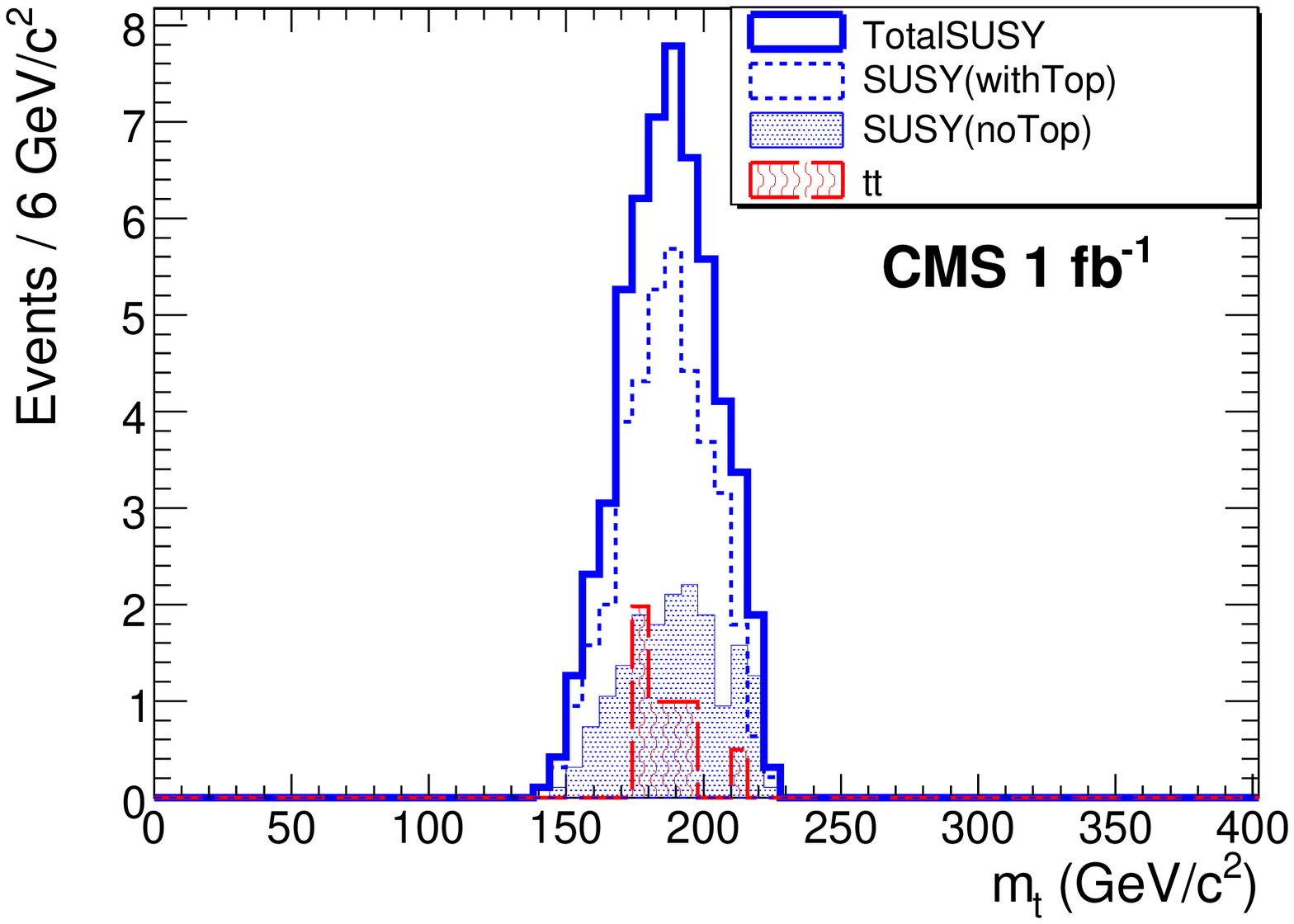}}
\caption{Missing transverse energy (left) and the extracted top
quark (right) after applying all cuts.}
  \label{fig.paktinat.METTop}
\end{center}
\end{figure}

After applying these cuts, the only remaining background is
$t\bar{t}$. The ratio of the SUSY signal against the SM
background is  11, when almost 70\% of the extracted SUSY events
have a generated top quark `SUSY(withTop)'.
 Figure \ref{fig.paktinat.METTop}
shows the distributions of missing transverse energy and the
extracted top quark for different samples. It can be seen that
the SUSY signal is well above the standard model ($t\bar{t}$)
background.  Including the systematic uncertainties on the
background \cite{BityukovCode} we estimate that the minimum
integrated luminosity for a 5$\sigma$ discovery is $\sim$0.25
$fb^{-1}$. Note that the analysis uses systematic uncertainties
that are realizable with 1 $fb^{-1}$ of data. For start-up (0.1
$fb^{-1}$) a separate study of the uncertainties needs to be
performed.

\section{CMS Reach in $m_0$-$m_{1/2}$ Plane}
\label{sec.reach} To estimate the reach over the mSUGRA parameter
space we apply the analysis selection path using the fast
simulation and reconstruction of CMS, FAMOS \cite{famos} and with
appropriate validation using the detailed simulation.

\begin{figure}[!Hhtb]
\begin{center}
  \includegraphics[width=60mm]{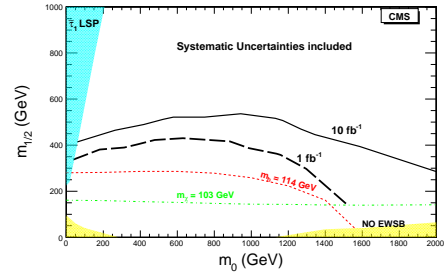}
\caption{The CMS experiment reach for mSUGRA in top+$\MET$ final
states in $m_0$-$m_{1/2}$ plane and for  $\tan\beta = 10$, $A_0 =
0$ and $\mu > 0$. The shaded regions are excluded because either
the $\sTau$ would be the LSP or because there is no radiative
electroweak symmetry breaking. The regions excluded by the LEP
limit on the $h^0$ or the $\chipm$ masses are delineated by
dashed lines.} \label{fig:ScanSys}
\end{center}
\end{figure}

The NLO cross section is calculated by PROSPINO \cite{prospino},
assuming only the $\sG-\sG$, $\sG-\sQ$ and $\sQ-\sQ$ productions
to be relevant.
%In the The minimum signal over background is as high as 40\%.
For 10 $fb^{-1}$ the jet energy scale and $b$-tagging
uncertainties expected are smaller compared to 1 $fb^{-1}$
\cite{PTDR-V1}. The jet energy scale relative uncertainty on the
final result  amounts to 11.3\%  while the relative systematic
uncertainty from the $b$-tagging is 7\% \cite{PTDR-V1}. The total
relative systematic uncertainty is 13.7\%. The reach result for 1
and 10 fb$^{-1}$ is shown in figure \ref{fig:ScanSys}.
The larger reach in the high $m_0$ region is due to the dominant
three body decay of the gluino to top quark in this region.

\section{Conclusion}
\label{Sec.Conclusion} We present the observability study of low
mass SUSY using an inclusive selection of events with a
hadronically decaying top quark  and large missing energy in the
final state. We estimate that for the test point LM1 the 5$\sigma$
discovery is achievable with  0.25 $fb^{-1}$. The CMS 5$\sigma$
reach contours in the mSUGRA parameter space for 1 and 10
 $fb^{-1}$  are also given.

\section{Acknowledgments}
I must thank my supervisors in this research, Dr. Luc Pape and
Dr. Maria Spiropulu for their kind help and support, also thanks
to the conference organizing committee specially Dr. Y. Farzan for
their help and hospitality. This work was done with the CMS
collaboration developed software.

\end{document}